\documentclass[11pt]{article}
\usepackage[dvips]{graphicx}
\usepackage{amsfonts}
\usepackage{amssymb}

\title{Net Fisher information measure versus ionization potential and dipole polarizability in atoms}
\author{K.~D.~Sen$^{1}$, C.~P.~Panos$^{2}$, K.~Ch.~Chatzisavvas$^{2}$\footnote{\texttt{e-mail:\, kchatz\,@\,auth.gr}}\,,
        \\ and Ch.~C.~Moustakidis$^{2}$
\\ \\
 {\it  $^{1}$School of Chemistry, University of Hyderabad},\\ {\it Hyderabad, 500046 India}\\
 {\it  $^{2}$Department of Theoretical Physics}, \\ {\it Aristotle University of Thessaloniki,}\\
                {\it  54124 Thessaloniki, Greece}
 }

\date{}

\begin{document}

\maketitle

\begin{abstract}
The net Fisher information measure $I_{T}$, defined as the product
of position and momentum Fisher information measures $I_{r}$ and
$I_{k}$ and derived from the non-relativistic Hartree-Fock wave
functions for atoms with $Z=1-102$, is found to correlate well
with the inverse of the experimental ionization potential. Strong
direct correlations of $I_{T}$ are also reported for the static
dipole polarizability of atoms with $Z=1-88$. The complexity
measure, defined as the ratio of the net Onicescu information
measure $E_{T}$ and $I_{T}$, exhibits clearly marked regions
corresponding to the periodicity of the atomic shell structure.
The reported correlations highlight the need for using the net
information measures in addition to either the position or
momentum space analogues. With reference to the correlation of the
experimental properties considered here, the net Fisher
information measure is found to be superior than the net Shannon
information entropy.
\end{abstract}

\begin{small}
\end{small}

\section{Introduction}

Two of the most popular information measures due to Shannon
\cite{Shannon48} and Fisher \cite{Fisher25} respectively, are
being increasingly applied in studying the electronic structure
and properties of atoms and molecules. The Shannon information
entropy $S_{r}$ of the electron density $\rho(\textbf{r})$ in
coordinate space is defined as
\begin{equation}\label{eq:eq1}
    S_{r}=-\int
    \rho(\textbf{r})\,\ln{\rho(\textbf{r})}\,d\textbf{r},
\end{equation}
and the corresponding momentum space entropy $S_{k}$ is given by
\begin{equation}\label{eq:eq2}
    S_{k}=-\int n(\textbf{k})\,\ln{n(\textbf{k})}\,d\textbf{k},
\end{equation}
where $n(\textbf{k})$ denotes the momentum density. The densities
$\rho(\textbf{r})$ and $n(\textbf{k})$ are respectively normalized
to unity and all quantities are given in atomic units. The Shannon
entropy sum $S_{T}=S_{r}+S_{k}$ contains the net information and
obeys the well known lower bound by Bialynicki-Birula and
Mycielski \cite{Bialynicki75} who obtained the entropic
uncertainty relation (EUR) which represents a stronger version of
the Heisenberg uncertainty principle of quantum mechanics.
Accordingly, the entropy sum in D-dimensions satisfies the
inequality \cite{Bialynicki75,Sears80}
\begin{equation}\label{eq:eq3}
    S_{T}=S_{r}+S_{k} \geq D\,(1+\ln{\pi}).
\end{equation}
Individual entropies $S_{r}$ and $S_{k}$ depend on the units used
to measure $r$ and $k$ respectively, but their sum $S_{T}$ does
not i.e. it is invariant to uniform scaling of coordinates.

The Shannon information entropies (uncertainty) provide a global
measure of information about the probability distribution in the
respective spaces. A more localized distribution in position space
corresponds to a \emph{smaller} value of information entropy. For
application of Shannon information entropy in chemical physics we
refer the reader to the published literature
\cite{Gadre100,Gadre101,Gadre102}. An example of quantification of
order of the chemical bonding employing Shannon information is
\cite{Karafiloglou04}.

Analogous applications for other quantum many-body systems
(nuclei, atomic clusters and correlated atoms in a trap-bosons)
have been reported recently \cite{Panos100}.

The Fisher information measure or intrinsic accuracy in position
space is defined as
\begin{equation}\label{eq:eq4}
    I_{r}=\int \frac{\left|\nabla\rho(\textbf{r})\right|^2}{\rho(\textbf{r})}
    \,d\textbf{r},
\end{equation}
and the corresponding momentum space measure is given by
\begin{equation}\label{eq:eq5}
    I_{k}=\int \frac{\left|\nabla n(\textbf{k})\right|^2}{n(\textbf{k})}
    \,d\textbf{k}.
\end{equation}

The individual Fisher measures are bounded through the Cramer-Rao
inequality \cite{Rao59} according to $\displaystyle{I_{r}\geq
\frac{1}{V_{r}}}$ and $\displaystyle{I_{k}\geq \frac{1}{V_{k}}}$,
where $V$'s denote the corresponding spatial and momentum
variances respectively. In position space, the Fisher information
measures the sharpness of probability density and for a Gaussian
distribution is exactly equal to the variance\cite{Frieden04}. A
sharp and strongly localized probability density gives rise to a
\emph{larger} value of Fisher information in the position space.
The Fisher measure in this sense is complementary to the Shannon
entropy and their \emph{reciprocal} proportionality is, in fact,
utilized in this work. The Fisher measure has the desirable
properties, i.e. it is always positive and reflects the
localization characteristics of the probability distribution more
sensitively than the Shannon information entropy \cite{Carroll06}.
However, for the electronic density distribution in atoms, the
enhanced sensitivity of the Fisher measure has not been
demonstrated explicitly. The lower bounds of Shannon sum
($S_{r}+S_{k}$) and Fisher product ($I_{r}I_{k}$) get saturated
for the Gaussian distributions. For a variety of applications of
the Fisher information measure we refer to the recent book
\cite{Frieden04} and for applications to the electronic structure
of atoms, to the pioneering work of Dehesa et al. \cite{Dehesa01}.

In the context of density functional theory (DFT), Sears, Parr and
Dinur \cite{Sears80b} were the first to highlight the importance
of Fisher information, by showing explicitly that the quantum
mechanical kinetic energy is a measure of the information content
of a distribution. A link of Shannon information entropy with the
kinetic energy for atomic clusters and nuclei has also been
indicated in \cite{Massen01}. The electron localization function
\cite{Becke01}, which has been widely successful in revealing the
localization properties of electron density in molecules, has been
interpreted in terms of Fisher information \cite{Roman01}.
Recently, the Euler equation of density functional theory has been
derived from the principle of the minimum Fisher information
within the time dependent versions \cite{Nagy03}. The Shannon
information sum $S_{T}$ has been used in a large majority of
applications of information theory in the electronic structure
studies involving atoms and molecules. In this work we define the
net information $I_{T}$ as the product $I_{r}I_{k}$ and consider
its inverse $I_{T}^{-1}$ as representing the net information
similar to $S_{T}$. In this sense we propose to employ $I_{T}$
instead of $S_{T}$ to assess the utility of the net Fisher
information vis-a-vis the Shannon entropy sum. This is done in the
analysis of experimental properties such as the ionization
potential and polarizability, corresponding to the neutral atoms
in their ground electronic states. It is worth noting here that
the net uncertainty measures defined in the conjugate spaces are
at the foundation of the quantum mechanical probability
distribution. As noted above, the quantities $I_{T}$ and $S_{T}$
measure the net information content of the probability
distribution including its spatial characteristics. Such measures
could therefore be tested in their ability to reproduce the trends
in atomic sizes, ionization potentials and the polarizabilities,
respectively.

Very recently the question whether atoms can grow in complexity
with the increase in nuclear charge has been addressed
\cite{Chatzisavvas05,Chatzisavvas06}. In particular the Onicescu
information measure \cite{Onicescu66} in position space $E_{r}$
and momentum $E_{k}$ have been defined as the corresponding
density expectation values $E_{r}=\int
\rho(\textbf{r})^2\,\d(\textbf{r})=\langle \rho(\textbf{r})
\rangle$ and  $E_{k}=\int n(\textbf{k})^2\,d(\textbf{k})=\langle
n(\textbf{k}) \rangle$, respectively.

The complexity $C$ is measured accordingly to the prescription due
to Lopez-Ruiz, Manchini and Calbet (LMC)
\cite{Lopez95,Chatzisavvas06} as
\begin{equation}\label{eq:eq6}
    C=S_{T} E_{T},
\end{equation}
where $E_{T}=E_{r}E_{k}$.

$S_{T}$ denotes the information content stored in the system and
$E_{T}$ corresponds to the disequilibrium of the system, i.e. the
distance from its actual state to equilibrium, according to
\cite{Lopez95}. Shiner, Davison and Landsberg (SDL)
\cite{Shiner99} and LMC measures were criticized in
\cite{Crutchfield00,Feldman98,Stoop05}. A related discussion can
be found in \cite{Chatzisavvas05,Chatzisavvas06}.

In the light of its sensitivity to describe the localization
property it is useful to consider in the above equation
$I_{T}^{-1}$ instead of $S_{T}$, to define LMC complexity measure
based on the net Fisher information. Thus a new definition of
complexity measure (LCM-like) is the following
\begin{equation}\label{eq:eq6b}
  C=E_{T} I_{T}^{-1}.
\end{equation}

It is generally agreed that the ionization potential (I.P.) and
static dipole polarizability $\alpha_{d}$ represent the two key
electronic properties of atoms and molecules, which control a host
of their other properties including chemical reactivity. Indeed,
the DFT descriptions of chemical reactivity \cite{Geerlings03}
such as electronegativity, hardness and Fukui functions are
intimately related to I.P. and $\alpha_{d}$. Due to this reason
several interesting studies have been made earlier
\cite{Fricke86,Politzer02} to find a correlation between the two
experimental properties. The purpose of this paper is to examine
how well the net Fisher information measure correlates with the
experimental values of the inverse of the first I.P. and
$\alpha_{d}$, of neutral atoms. In particular, we shall be
interested in the relative advantages of using the net Fisher
information over the Shannon information entropy sum. In a similar
manner we also report here the LMC complexity measure based on
$I_{T}^{-1}$ and point out some of the new features which are not
displayed when the $S_{T}$ is used instead \cite{Chatzisavvas06}.

\section{Results}
\subsection{Ionization potentials and $I_{T}$}\label{sub:sub1}

Throughout this work we have used the consistent data on the
\emph{spherically} averaged density \cite{Romera02} derived from
the highly accurate analytic Hartree-Fock (HF) wave functions
\cite{Koga02}. These results are in quantitative agreement with
our results reported earlier, i.e. those derived form the analytic
HF wave functions due to Bunge et al \cite{Bunge93}. In a large
number of cases, we have also computed the information measures
using numerical HF wave functions and have found similar
quantitative agreement. We note here that the present results do
not include any electron correlation and/or relativistic effects.
The experimental ionization potential of neutral atoms has been
taken from reference \cite{Parr82}.

\begin{figure}
 \centering
 \includegraphics[height=10.0cm,width=10.0cm]{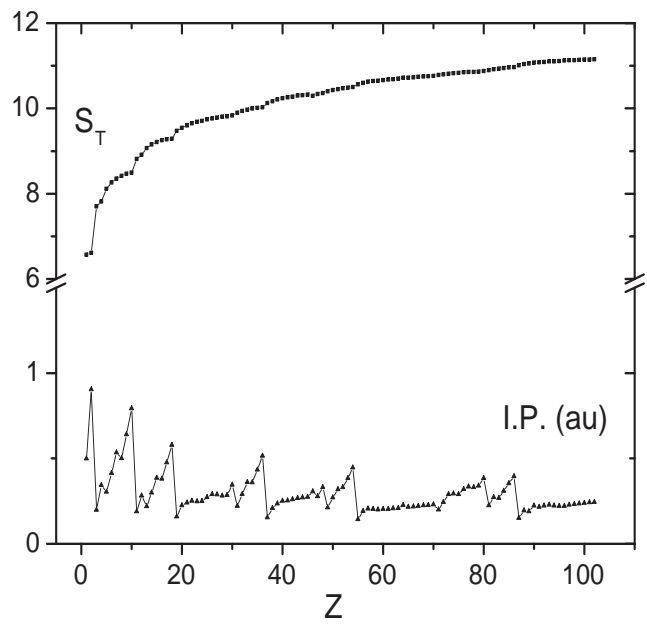}
 \caption{$S_{T}$ and I.P. as functions of $Z$ \label{fig:1}}
\end{figure}

In Fig. \ref{fig:1} we plot $S_{T}$ and I.P. as functions of $Z$.
It is observed that $S_{T}$ does not perform as a sensitive
information measure reproducing the details of the trends in I.P.
It does show the gross atomic periodicity in terms of the shell
structure as the humps.

\begin{figure}
 \centering
 \includegraphics[height=10.0cm,width=10.0cm]{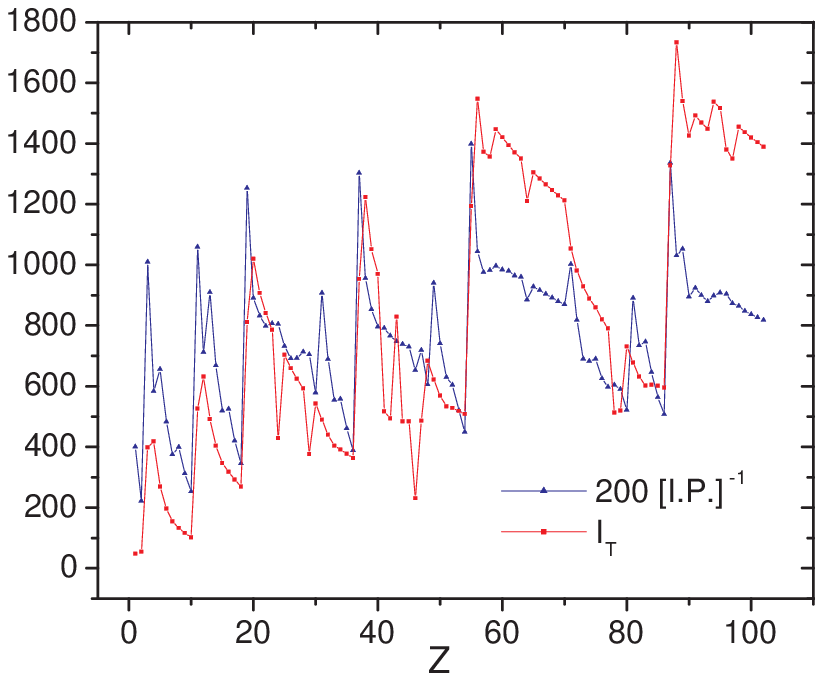}
 \caption{$I_{T}$ and I.P. as functions of $Z$ \label{fig:2}}
\end{figure}

In Fig. \ref{fig:2} the values of $I_{T}$ (instead of $S_{T}$ as
in Fig. \ref{fig:1} and the \emph{inverse} of I.P. are plotted as
functions of $Z$. Compared to Fig. \ref{fig:1} the two curves in
Fig. \ref{fig:2}, resemble each other in far more details. It has
been shown earlier that $I_{k}$ behaves similarly to I.P. for
atoms \cite{Romera02}. The net Fisher information amplifies the
details of correlation by approximately two orders of magnitude.
Our aim of plotting $I_{T}$ versus inverse of I.P. is to show the
similarities in the two curves in the \emph{upward} direction and
also lay emphasis on the net Fisher information, $I_{T}$. We note
here that in subsection \ref{sub:sub3} it is $I_{T}^{-1}$ that
enters the definition of complexity which directly correlates with
I.P. Very recently, the idea of taking the relative Shannon
entropy of an element within a group of the periodic table with
respect to the inert gas atom, located at the end of the group has
been proposed \cite{Borgoo04} in order to get a more sensitive
quantum similarity measure of density distributions. It would be
interesting to investigate whether the correlations found in Fig.
\ref{fig:2}) can be further improved by using such a similarity
measure for each group using the Fisher information according to
\begin{equation}
  \Omega(Z)=1-\left[ \frac{I_{T(ref)}}{I_{T(Z)}} \right],
\end{equation}
where $\Omega(Z)$ measures the distance in compactness of the
element $Z$ from the most compact ideal gas atom in the same
group, used as reference. A larger value of $\Omega(Z)$ would
correspond to smaller I.P. Our use of $\Omega(Z)$ is inspired by
Landsberg's definition of order parameter $\Omega:
\Omega=1-\Delta$, where $\displaystyle{\Delta=\frac{S_{T}}{S_{\rm
max}}}$ is a disorder parameter given in terms of the actual
entropy $S$ and the maximum possible entropy $S_{\rm max}$ of the
system \cite{Landsberg84}. An application of $\Omega$ in quantum
many-body systems was carried out in \cite{Panos01}.

\begin{figure}
 \centering
 \includegraphics[height=10.0cm,width=10.0cm]{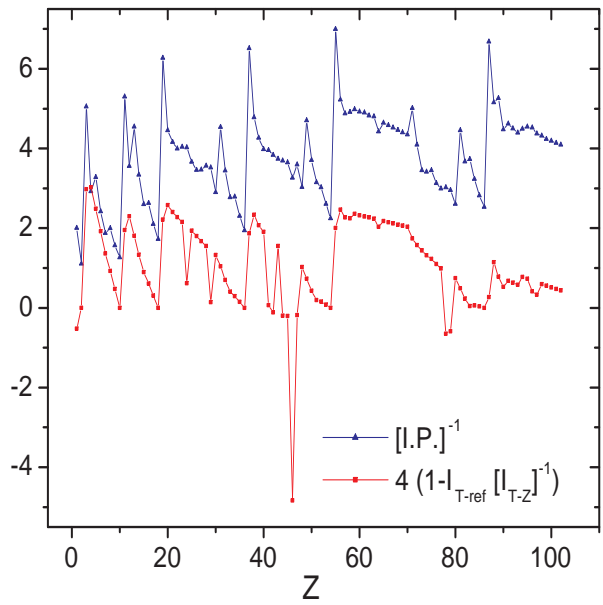}
 \caption{$\Omega(Z)$ and [I.P.]$^{-1}$ as functions of
$Z$ \label{fig:3}}
\end{figure}

In Fig. \ref{fig:3} we present $\Omega$ and [I.P.]$^{-1}$ as
functions of $Z$. It is found that the correlation is more direct
than that obtained in Fig. \ref{fig:2}. This observation suggests
that $\Omega(Z)$ can be used as a measure of quantum similarity of
atoms and opens up a new application of the net Fisher information
measure $I_{T}$.

\subsection{Dipole polarizability and $I_{T}$}\label{sub:sub2}

\begin{figure}
 \centering
 \includegraphics[height=10.0cm,width=10.0cm]{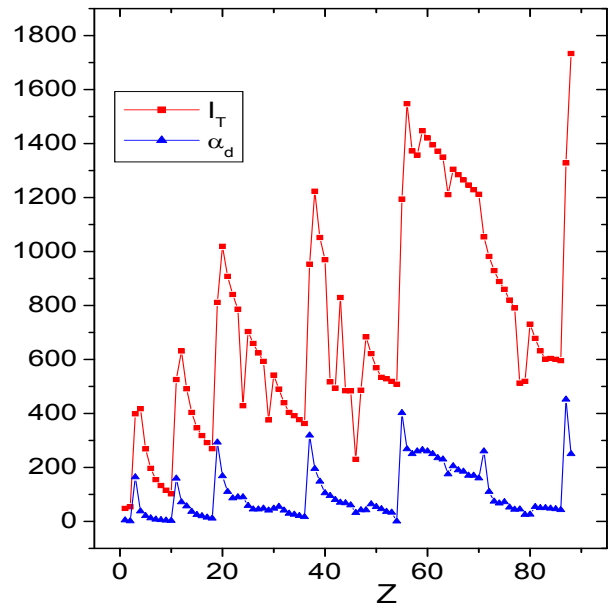}
 \caption{$I_{T}$ and $\alpha_{d}$ as functions of
$Z$ \label{fig:4}}
\end{figure}

The variation of polarizability $\alpha_{d}$ of atoms with $S_{T}$
is found to be essentially similar to that of $S_{T}$ versus I.P.,
as already given in Fig. \ref{fig:1}. In the background of such
insensitivity of $S_{T}$, we shall now consider the correlation of
$I_{T}$ with the experimental estimates of polarizability
$\alpha_{d}$. The experimental values have been taken from the
compilation of Miller and Bederson \cite{Miller77} for atoms with
$Z=1-88$. The variation of $I_{T}$ and $\alpha_{d}$ with $Z$ for
atoms with $Z=1-88$ is shown in Fig. \ref{fig:4}. The overall
correlation is found to be excellent with the maximum
polarizability elements of the alkali atoms immediately following
the sharply increasing values if $I_{T}$ just after the inert gas
atoms. The polarizability predicted by $I_{T}$ for the alkaline
atoms present themselves as the only examples which are not
sufficiently well discriminated against the neighboring atoms, in
this case, the alkali metal atoms. It appears that the compactness
described by $I_{T}$ in going from the valence electron
configuration of $(ns)^1$ to $(ns)^2$ does not increase sharply
enough to quantitatively reflect the changes in polarizability
from alkali to alkaline earth atoms. For these examples it is
advisable to carry out further computations of the Fisher
information using wave functions which include electron
correlation effects.

\subsection{Complexity using $I_{T}$}\label{sub:sub3}

\begin{figure}
 \centering
 \includegraphics[height=10.0cm,width=10.0cm]{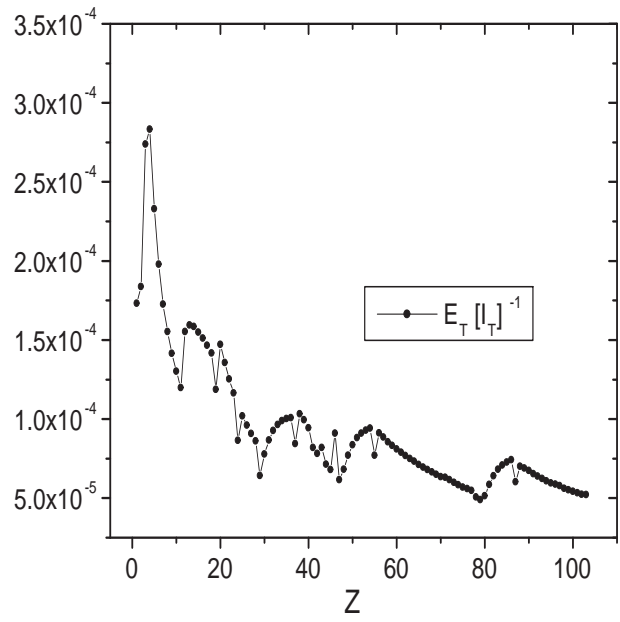}
 \caption{$C=E_{T}I_{T}^{-1}$ as a function of $Z$ \label{fig:5}}
\end{figure}

\begin{figure}
 \centering
 \includegraphics[height=10.0cm,width=10.0cm]{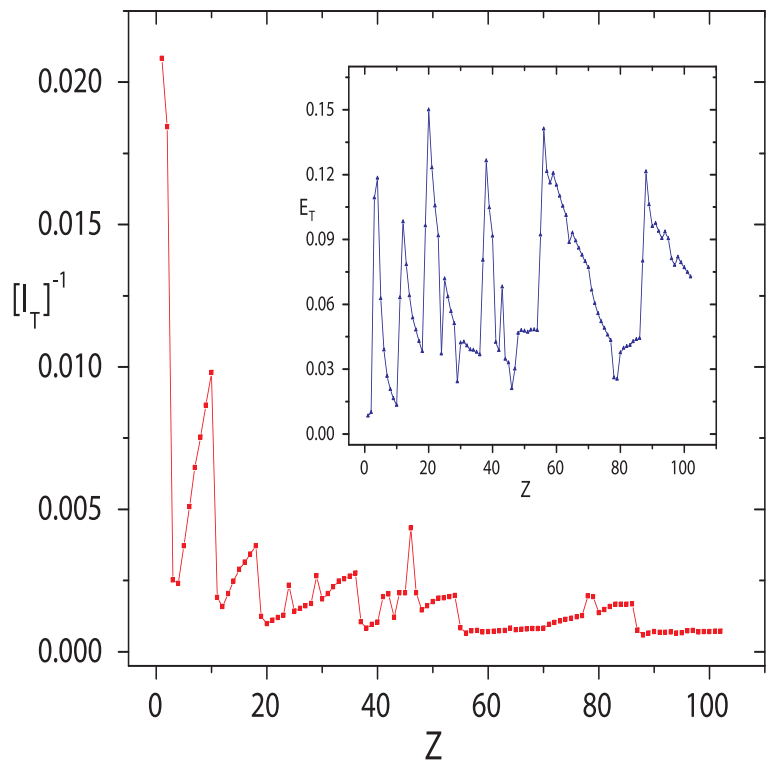}
 \caption{$E_{T}$ and $I_{T}^{-1}$ as functions of $Z$ \label{fig:6}}
\end{figure}

Finally, we discuss the LMC measure of complexity using
Eq.(\ref{eq:eq6}) in which $S_{T}$ is substituted with
$I_{T}^{-1}$. In an earlier publication \cite{Chatzisavvas05}, the
SDL measure of complexity (for various indices of disorder and
order) has been plotted as a function of $Z$ and a series of
oscillations around a certain average value, was obtained. This
led to the conclusion that atoms cannot grow in complexity as $Z$
increases. The latter conclusion was modified in
\cite{Chatzisavvas06}, where a similarity of SDL (for magnitude of
disorder equal to zero and magnitude of order equal to four) and
LMC measures led to the observation that $C$ is an increasing
function of $Z$. Here, in Fig. \ref{fig:5}, we present the measure
$C=E_{T} I_{T}^{-1}$ as a function of $Z$. This curve points to a
gradual decrease in complexity with systematic oscillations due to
new shells added on as $Z$ increases. It is seen that while the
fluctuations of complexity following the periodicity of the
elements represent a general feature, the issue of the behavior of
complexity (increase or decrease) with increasing $Z$, cannot be
answered at present with certainty. Such an answer, would be
desirable towards the proper description of organization of
quantum systems. A separate plot of $E_{T}$ and $I_{T}^{-1}$ in
Fig. \ref{fig:6} suggests that the trend shown in Fig. \ref{fig:5}
is controlled by $I_{T}^{-1}$. While the earlier conclusion about
the oscillations of complexity in atoms stands vindicated, the use
of Fisher information measure leads to a more transparent
variation of $C$ as a function of $Z$. It is also indicated that
the decreasing complexity with increasing $Z$ is bounded by the
successive shells as oscillations.

\section{Summary}

In conclusion, we have found that the \emph{net} Fisher
information denoted by the product of the Fisher information in
position and momentum spaces (Eq.(\ref{eq:eq4})-(\ref{eq:eq5}))
describes the variation of the I.P. and the static dipole
polarizability as a function of $Z$, more efficiently than the
Shannon information entropy sum
(Eq.(\ref{eq:eq1})-(\ref{eq:eq2})). Our results also highlight the
importance of using the net information in addition to those
corresponding to either the position or momentum space separately,
in order to analyze and predict the experimental properties.
Furthermore, the LMC measure of complexity of atoms as a function
of $Z$ is found that can be transparently expressed employing the
net Fisher information. It would be interesting to extend the
application of net Fisher information entropy to nuclear,
molecular and atomic cluster densities.

\section*{Acknowledgements}

K.D. Sen acknowledges with thanks the Department of Theoretical
Physics, Aristotle University of Thessaloniki, for its warm
hospitality. He is also grateful to Prof. Antoniou Ioannis for
constant encouragement and to Professor E. Romera for data
sharing. The work of K.D. Sen, C.P. Panos and K.Ch. Chatzisavvas,
was supported by Herakleitos Research Scholarships (21866) of
$\textrm{E}\Pi\textrm{EAEK}$ and the European Union.


\clearpage
\newpage

\clearpage
\newpage

\clearpage
\newpage

\clearpage
\newpage

\clearpage
\newpage

\clearpage
\newpage

\end{document}